\documentclass{birkjour}
 \newtheorem{thm}{Theorem}[section]

 \theoremstyle{definition}
 \newtheorem{defn}[thm]{Definition}
 \theoremstyle{remark}

 \numberwithin{equation}{section}

\newcommand{\eq}[1]{\begin{equation}
                     \begin{split} #1 \end{split}
                     \end{equation}}

\newcommand{\alg}{T^*\Pi A}

\begin{document}

%
%
%
%
%
%
%
%
%

\title[Star products and $\alpha'$-corrections]
 {Star products on graded manifolds and \\ $\alpha'$-corrections to double field theory}

\author[Andreas Deser]{Andreas Deser}

\address{%
Appelstr.2\\
D-30167 Hannover}

\email{andreas.deser@itp.uni-hannover.de}




\date{September 20, 2015}

\begin{abstract}
Originally proposed as an $O(d,d)$-invariant formulation of classical closed string theory, double field theory (DFT) offers a rich source of mathematical structures. Most prominently, its gauge algebra is determined by the so-called C-bracket, a generalization of the Courant bracket of generalized geometry, in the sense that it reduces to the latter by restricting the theory to solutions of a ``strong constraint''. Recently, infinitesimal deformations of these structures in the string sigma model coupling $\alpha'$ were found. In this short contribution, we review constructing the Drinfel'd double of a Lie bialgebroid and offer how this can be applied to reproduce the C-bracket of DFT in terms of Poisson brackets. As a consequence, we are able to explain the $\alpha'$-deformations via a graded version of the Moyal-Weyl product in a class of examples.
We conclude with comments on the relation between $B$- and $\beta$-transformations in generalized geometry and the Atiyah algebra on the Drinfel'd double.
\end{abstract}

\maketitle
\section{Introduction: Brackets and deformations in DFT}
Due to the extended nature of closed strings moving in a background, the field theory describing its classical dynamics is different from that of a point particle. In particular, the string can ``wind'' around compact cycles of the background manifold. This gives rise to two sets of parameters (or zero modes) characterizing the solutions to the classical equations of motion. One of them is associated to the center of mass momentum $p_i$ of the closed string and the corresponding configuration space coordinates $x^i$ span the phase space of the center of mass treated as a point particle. The second set $\tilde p^i$ is associated to the winding and gives rise to a second set of coordinates $\tilde x_i$. DFT is a field theory on this ``doubled configuration space'' which can be reduced to ordinary configuration space by using the strong constraint
\eq{
\partial_i\phi(x,\tilde x)\,\tilde \partial^i \psi(x,\tilde x) + \tilde \partial^i\phi(x,\tilde x)\,\partial_i\psi(x,\tilde x) = \;0, 
}
for functions $\phi,\psi$ on the doubled configuration space. This constraint has its origin in the level matching condition for physical fields in string theory and restores the right amount of coordinates of a physical configuration space. We refer to the reader especially to \cite{Hull:2009mi} and the lecture notes \cite{Zwiebachlectures} for an introduction to DFT.  

\subsection{C-bracket and bilinear form}
In \cite{Hull:2009zb, Hohm:2010jy}, a Lagrangian action for DFT was formulated and gauge symmetries were identified. Due to a lack of space, we only present results that are important for the rest of the presentation. To state the gauge symmetries, we use notation conventions of \emph{generalized geometry}. On a $d$-dimensional manifold $M$, generalized vector fields $V$ are locally given by sections of $TM\oplus T^*M$, i.e. $V=V^i\partial_i + V_idx^i$. To state local expressions in DFT, the components are allowed to depend on the doubled configuration space with coordinates $(x^i,\tilde x_i)$. Furthermore one uses a capital index to denote objects transforming in the fundamental representation of $O(d,d)$, i.e. $V^M=(V^i(x,\tilde x),V_i(x,\tilde x))$, where $A\in O(d,d)$ obeys
\eq{
A \eta A^t =\;\eta\;, \qquad \eta = \begin{pmatrix} 
0 & id \\
id & 0 
\end{pmatrix}\;,
}
and $id$ is the d-dimensional identity matrix. We will denote the bilinear form represented by $\eta$ by $\langle \cdot,\cdot\rangle$. Capital indices are raised and lowered by the latter, so for generalized vectors $V,W$ we have 
\eq{
\langle V,W\rangle =\;V^PW^Q \eta_{PQ} =\; V^iW_i + V_i W^i\;.
}
The gauge symmetries of DFT are given by the action of a generalized Lie derivative, acting on functions $\phi$ by\footnote{We use the notation $\partial_M$ for the pair $(\partial_i,\tilde \partial^i)$, so expressions like $V^M\partial_M$ are expanded as $V^i\partial_i + V_i \tilde \partial^i$.} $\mathcal{L}_V \phi = V^K\partial_K \phi$ and generalized vectors $W$ according to
\eq{
(\mathcal{L}_V W)_K =\;&V^P\partial_P W_K + (\partial_K V^P - \partial^P V_K) W_P\;,\\
(\mathcal{L}_V W)^K =\;&V^P\partial_P W^K -(\partial_P V^K - \partial^K V_P)W^P\;.
}
Finally, the commutator of two generalized Lie derivative gives the generalized Lie derivative with respect to the \emph{C-bracket} of two generalized vectors, which is given in components by
\eq{
\label{Cbracket}
\Bigl([V,W]_C\Bigr)^P =\;V^K\partial_K W^P - W^K\partial_K V^P -\frac{1}{2}\Bigl(V^K\partial^P W_K - W^K\partial^PV_K\Bigr)\;.
}
Note that for the specific solution $\tilde \partial^i=0$, this bracket reduces to the well-known Courant bracket of generalized geometry. In the following subsection, we will present a deformation of the bilinear form $\eta$ and the C-bracket found in double field theory.   

\subsection{$\alpha'$-deformations}
Classical closed string theory is described by a two-dimensional sigma model. Perturbative expansions are formal power series in the coupling constant $\alpha' = l_s^2$, where $l_s$ is the fundamental string length. Recently, corrections to the bilinear form and C-bracket up to first order in $\alpha'$ were given in \cite{Hohm:2013jaa, Hohm:2014eba}. For the correction of the bilinear form, we introduce the notation $\langle V,W\rangle_{\alpha'}:=\;\langle V,W\rangle - \alpha'\langle\langle V,W\rangle\rangle$, where the component expression for the correction is 
\eq{
\label{bilinearcorr}
\langle\langle V, W\rangle \rangle =\;\partial_P V^Q \partial_Q W^P\;.
}
Similarly, for the correction to the C-bracket, we introduce the short notation $[V,W]_{\alpha'} :=\; [V,W]_C - \alpha'[[V,W]]$, where the correction is given by
\eq{
[[V,W]]^K = \;\frac{1}{2}\Bigl(\partial^K\partial_Q V^P \partial_P W^Q - V\leftrightarrow W\Bigr)\;.
}
Note that this expression has a form part and a vector part. As an example, we expand the vector part in terms of partial derivatives:
\eq{
\label{cbracketcorr}
[[V,W]]_i =\;\frac{1}{2}\Bigl(&\partial_i\partial_m V^n\partial_n W^m + \partial_i \partial_m V_n \tilde \partial^n W^m + \partial_i\tilde \partial^m V^n\partial_n W_m \\
&+\partial_i\tilde \partial^m V_n \tilde \partial^n W_m - V\leftrightarrow W\Bigr)\;.
}
The goal of this work is to get a systematic explanation of the derivative expansions \eqref{bilinearcorr} and \eqref{cbracketcorr}.  In the following section, we are going to set up a mathematical formalism to rewrite the bilinear form and the C-bracket in terms of Poisson brackets. This will allow us finally to identify the deformation using a Moyal-Weyl star product on a specific symplectic supermanifold. 

\section{Lie bialgebroids and double fields}
For finite dimensional vector spaces $\mathcal{V}$, it is a standard exercise to show the isomorphism between the exterior algebra and the algebra of polynomials in the parity reversed version $\Pi\mathcal{V}$:
\eq{
\label{pi}
\wedge^\bullet \mathcal{V}^* \simeq \textrm{Pol}^\bullet(\Pi \mathcal{V})\;.
}
For a finite dimensional $\mathbb{Z}_2$-graded vector space $\mathcal{W} = \mathcal{W}_0 \oplus \mathcal{W}_1$, parity reversion $\Pi$ acts according to $(\Pi \mathcal{W})_0 = \mathcal{W}_1$ and $(\Pi \mathcal{W})_1 = \mathcal{W}_0$. In \eqref{pi}, elements of $\mathcal{V}$ have degree $0$ and elements of $\Pi\mathcal{V}$ have degree $1$. In the case of vector bundles, differentials are derivations of the exterior algebra, which get mapped to derivations on functions, i.e. vector fields. Squaring to zero means that the vector fields are actually \emph{homological}. These statements are summarized by the structure of a Lie algebroid:
\begin{defn}
A \emph{Lie algebroid} is a vector bundle $A\rightarrow M$ together with a homological vector field $d_A$ of degree 1 on the supermanifold $\Pi A$.
\end{defn}
A pair $(A,A^*)$ of a Lie algebroid and its linear dual has the structure of a \emph{Lie bialgebroid} if the differentials respect the brackets on the dual spaces. This will be the basic structure used in the following sections. 

\subsection{Lie bialgebroids and the Drinfel'd double}
Let $(A,A^*)$ be a pair of dual Lie algebroids over a manifold $M$. The homological vector field $d_A$ can be lifted to a function on the cotangent bundle $T^*\Pi A \overset{\mathfrak{p}}{\rightarrow} \Pi A$. Similarly, the corresponding operator $d_{A^*}$ for the dual can be lifted to $T^*\Pi A^* \overset{\bar{\mathfrak{p}}}{\rightarrow} \Pi A^*$. Similarly to the case of standard phase spaces, there is a Legendre transform $L: \alg \rightarrow T^*\Pi A^*$, which can be used to pull back functions. Thus we have the situation
\eq{
\label{diagram}
\begin{matrix}
T^*\Pi A & \overset{L}\rightarrow & T^*\Pi A^* \\
\downarrow \mathfrak{p} & & \downarrow \bar{\mathfrak{p}} \\
\Pi A & & \Pi A^* 
\end{matrix}
}
For local formulas we use coordinates $(x^i,\xi^a)$ on $\Pi A$, where $x^i$ are coordinates on the base manifold and $\xi^a$ denote the (Grassmann odd) fibre coordinates. On its cotangent bundle, we have in addition the canonical conjugate momenta, i.e. $(x^i,\xi^a,x^*_i, \xi^*_a)$. As in the purely even case, there is a canonical Poisson bracket on $\alg$, given by the relations
\eq{
\lbrace x^i,x^*_j\rbrace = \; \delta^i_j\;,\qquad \lbrace \xi^a,\xi^*_b\rbrace =\; \delta^a_b\;.
}
Using this Poisson structure and the ``lifted'' vector field 
\eq{
\label{theta}
\theta :=\; h_{d_A} + L^*h_{d_{A^*}}\;,
}
it is possible to write down the following concise characterization of $(A,A^*)$ being a Lie bialgebroid:
\begin{thm}
\label{Bialgthm}
The pair $(A,A^*)$ is a Lie bialgebroid if and only if $\lbrace \theta, \theta\} =0\;.$
\end{thm}
We refer to \cite{Deethesis} for a proof and further details on the mathematical structures introduced in the present work. Theorem \ref{Bialgthm} is the motivation for the following definition:
\begin{defn}
For a Lie bialgebroid as above, the bundle $\alg$, equipped with the homological vector field $\lbrace \theta, \cdot\rbrace$ is called the \emph{Drinfel'd double} of $(A,A^*)$.
\end{defn}
We refer to \cite{Mack1, Mack2} for the original work on the Drinfel'd double in this context. The essential ingredient for the homological vector field is the function $\theta$ in \eqref{theta}.

\subsection{C-bracket in terms of Poisson brackets}
Let $M$ be a Poisson manifold. Then the standard example of a Lie bialgebroid is $(A,A^*)=(TM,T^*M)$. The respective brackets are the Lie bracket and Koszul bracket\footnote{The Koszul bracket of forms $\omega_1,\omega_2 \in \Gamma(T^*M)$ is given by 
$$[\omega_1,\omega_2]_K = \mathcal{L}_{\pi^\sharp(\omega_1)}\omega_2 - \iota_{\pi^\sharp(\omega_2)}d\omega_1\;,$$ where $\mathcal{L}$ is the Lie derivative and $\pi^\sharp$ is the anchor determined by the Poisson structure.}, giving rise to the de Rham and Poisson-Lichnerowicz differential, respectively. We use their lifts to functions on the Drinfel'd double to define two sets of momentum variables $p_i,\tilde p^i$:
\eq{
h_{d_A} &=\;a^j_i(x)x^*_j\xi^i -\tfrac{1}{2}f^k_{ij}(x)\xi^i\xi^j\xi^*_k =:\; \xi^ip_i\;,\\
h_{d_{A^*}} &=\;a^{ij}(x)x^*_i\xi^*_j + \tfrac{1}{2}Q_k^{ij}(x)\xi^k\xi^*_i\xi^*_j =:\; \xi^*_i \tilde p^i\;, 
}
where we denote the anchor maps by $a^j_i$ and $a^{ij}$, and $f$ and $Q$ are determined by the brackets on $A$ and $A^*$, respectively\footnote{The notation $f$ and $Q$ is common in the physics literature, where these objects play a role in flux compactifications of string theory.}. We consider the momenta $p_i$ and $\tilde p^i$ to act on functions on $\alg$ by using the Poisson bracket, e.g. $\lbrace p_i, \cdot\rbrace$. In particular, lifting functions $\phi \in \mathcal{C}^\infty(M)$ to $\alg$ (we use the same letter $\phi$ for the lift), we define the following two differential operators: 
\eq{
\partial_i \phi :=\;\lbrace p_i, \phi\rbrace\;,\qquad \tilde \partial^i \phi :=\; \lbrace \tilde p^i,\phi\rbrace\;.
}
Lifting furthermore generalized vectors to $\alg$, i.e. if locally $X^i\partial_i + \omega_idx^i \in \Gamma(TM\oplus T^*M)$, we define $V:= X^i\xi^*_i + \omega_i \xi^i \in \alg$, we are able to show the following result by rewriting the proof done in \cite{Deethesis} for Courant brackets, but using $\partial_i$ and $\tilde \partial^i$ here:
\begin{thm}
\label{thm1}
For vanishing $f$ and $Q$, let $V,W$ be lifts of generalized vectors to $\alg$. Furthermore, define the Dorfmann-product $\circ$ by
\eq{
V\circ W :=\; \Bigl\lbrace \lbrace \xi^i p_i + \xi^*_i \tilde p^i, V\rbrace,W\Bigr\rbrace\;.
}
Then the C-bracket of $V,W$ (lifted to $\alg$) is given by
\eq{
[V,W]_C =\; \frac{1}{2}\Bigl(V\circ W - W\circ V\Bigr)\;.
}
\end{thm}
The proof is an easy evaluation in local coordinates of $\alg$, and comparison with \eqref{Cbracket}, see \cite{Deser:2014mxa}. The generalization for non-vanishing $f$ and $Q$ would give a version of the C-bracket containing ``fluxes'', which, as far as we know, has not been done so far in the physics literature. As a final remark for this subsection, we observe that the bilinear form $\langle V,W\rangle$ is given by evaluating the Poisson bracket $\lbrace V,W\rbrace$ of the lifted quantities to $\alg$. These observations will be used in the following sections to suggest a way to understand the deformations \eqref{bilinearcorr} and \eqref{cbracketcorr} of the bilinear form and C-bracket encountered in DFT.

\section{Deformation of the metric and C-bracket}
The result of theorem \ref{thm1} immediately suggests the interpretation of $\alpha'$-corrections such as \eqref{bilinearcorr} and \eqref{cbracketcorr} in terms of deformation theory. Given a formal star product on the algebra of smooth functions on a Poisson manifold\footnote{More precisely on formal power series in a deformation parameter $t$, usually denoted by ${\mathcal C}^\infty(M)[[t]]$. We refer to \cite{Blumenhagen:2011ph, Blumenhagen:2013zpa, Bakas:2013jwa, Blumenhagen:2014sba} for recent applications of deformation theory in closed string theory and to \cite{Bordemann:1999ca, Klemm:2001yu, KellerWaldmann} for star products on graded manifolds.}, the star-commutator reproduces the Poisson bracket in the first non-trivial order:
\eq{
\lbrace f,g\rbrace =\;\underset{t\rightarrow \infty}{\lim}\frac{1}{t}\Bigl(f\star g - g\star f\Bigr)\;.
}
Thus, higher orders lead to deformations of the Poisson bracket and as a consequence of theorem \ref{thm1} of the metric and C-bracket. In the following, we will define an appropriate notion of star-commutator taking into account the Koszul signs on the graded manifold $\alg$. Furthermore, we will give a (constant) Poisson structure on $\alg$ such that the corrections of DFT are reproduced by taking star-commutators w.r.t. the corresponding Moyal-Weyl product. 

\subsection{Star-commutator and Poisson structure}
For the Moyal-Weyl case, let $I=i_1\cdots i_k,\;J=j_1\cdots j_k$, with $\partial_I = \partial_{x^{i_1}}\cdots \partial_{x^{i_k}}$, then the star commutator for purely even manifolds has the standard form
\eq{
\lbrace f, g\rbrace^* = \; \overset{\infty}{\underset{k=1}{\sum}}\,t^k\Bigl(\underset{IJ}{\sum}\,m_k^{IJ}(\partial_I f \partial_J g - \partial_I g \partial_J f)\Bigr)\;.
}
In the case of the symplectic supermanifold $\alg$, we will replace this by the following expression:
\eq{
\label{supercomm}
\lbrace f,g\rbrace^* =\; \overset{\infty}{\underset{k=1}{\sum}}\;t^k\Bigl(\underset{IJ}{\sum}\,(\partial_I f \partial_J g - (-1)^\epsilon \partial_I g \partial_J f )\Bigr)\;.
}
The sign $(-1)^\epsilon$ takes care of the $\mathbb{Z}_2$-grading and is given by
\eq{
\epsilon =\; \lvert f\rvert \lvert g \rvert + \lvert x^J \rvert(\lvert f \rvert -1) +\lvert x^I \rvert(\lvert g \rvert -1)\;,
}
where $\lvert f \rvert$ denotes the $\mathbb{Z}_2$-degree of a function and the shorthand notation $\lvert x^I\rvert := \lvert x^{i_1}\rvert + \dots + \lvert x^{i_k}\rvert$ is used. We remark that in contrast to the Moyal-Weyl case where the odd powers of the deformation don't contribute due to the antisymmetry of the Poisson tensor, in the graded case there are such contributions due to the different sign rule. In our case this will open the possibility to get the appropriate $\alpha'$-correction. 

Finally, we have to choose a Poisson structure on $\alg$ which correctly reproduces both, the correction to the bilinear form $\langle \cdot,\cdot \rangle$ and the C-bracket. Furthermore, the corresponding Poisson brackets, i.e. the first order star commutators still have to give the result of theorem \ref{thm1}. It turns out that this is indeed possible. To avoid long calculations we choose a setup which is as simple as possible, but still shows the essential features. Let $M$ be a symplectic manifold with Poisson tensor $\pi$. In this case $(TM,T^*M)$ is a Lie bialgebroid. In the expressions for the $\alpha'$-corrections, there are no $f$ -- and $Q$ -- fluxes. We can achieve the latter by taking the standard basis of vector fields on the tangent bundle. As a consequence, we get
\eq{
h_{d_A} =\; \xi^m x^*_m\;, \qquad L^*h_{d_{A^*}} =\; \xi^*_m \pi^{mn} x^*_n\;.
}
This is a special solution to the strong constraint of double field theory, with $\tilde{\partial}^i f = \lbrace \tilde p^i,f\rbrace = \pi^{ij}\partial_j f$. We choose the following Poisson structure on the Drinfel'd double:
\eq{
\label{specialPoisson}
\pi_{\alg}=\;\frac{\partial}{\partial x^*_i} \wedge \frac{\partial}{\partial x^i} + \frac{\partial}{\partial \xi^*_i}\wedge \frac{\partial}{\partial \xi^i} + \frac{\partial}{\partial x^i}\wedge\frac{\partial}{\partial \xi^*_i} -\pi^{ij}\frac{\partial}{\partial x^i}\wedge \frac{\partial}{\partial \xi^j}\;.
}
We will give our results for the deformation for this situation. In the general case, we have a differential operator $\tilde \partial^i = \lbrace p^i,\cdot\rbrace$, whose action on functions depends on the chosen Lie bialgebroid. If it is possible to associate a vector field $\tfrac{\partial}{\partial \tilde x_i}$ to this operator, the corresponding Poisson structure would be
\eq{
\pi_{\alg} =\; \frac{\partial}{\partial x^*_i} \wedge \frac{\partial}{\partial x^i} + \frac{\partial}{\partial \xi^*_i}\wedge \frac{\partial}{\partial \xi^i} + \frac{\partial}{\partial x^i}\wedge\frac{\partial}{\partial \xi^*_i} + \frac{\partial}{\partial \tilde x_i} \wedge \frac{\partial}{\partial \xi^i}\;.
}
We will leave the investigation of existence and properties of such a Poisson structure and its relation to double field theory for future work and give our deformation results for the Poisson tensor \eqref{specialPoisson} in the following. 

\subsection{Deformation of the metric}
Due to the various terms of the graded Poisson structure \eqref{specialPoisson}, computing higher orders of the graded Moyal--Weyl product is lengthy, but straight forward. We therefore refer the reader to \cite{Deser:2014wva} for computational details and only give the results. We will use the notation $\tilde \partial^i$ for $\lbrace p^i,\cdot\}$. Furthermore, we use the following notation for star -- commutators:
\eq{
\lbrace f, g\rbrace =\; \overset{\infty}{\underset{k=1}{\sum}}\; t^k\lbrace f,g\rbrace_{(k)}\;.
}
Taking $V = V^m(x)\xi^*_m + V_m(x)\xi^m$ and $W = W^m(x)\xi^*_m + W_m(x)\xi^m$ to be lifts of generalized vectors to $\alg$, we get the following results for the first two orders in the deformation parameter:
\eq{
\lbrace V, W\rbrace_{(1)} &=\, (V^i W_i + V_i W^i) =\,\langle V,W\rangle\;, \\
 \lbrace V, W\rbrace_{(2)}&=\, - \partial_i V^j\partial_j W^i - \partial_i V_j \tilde \partial^j W^i -\tilde\partial^i V^j \partial_j W_i - \tilde \partial^i V_j \tilde \partial^j W_i\;.
}
Comparing the latter expressions with the formulas from DFT \eqref{bilinearcorr}, we get the following statement:
\begin{thm}
Let $V=V^i\xi^*_i + V_i \xi^i$ and $W=W^i\xi^*_i + W_i\xi^i$ be two generalized vectors, lifted to $\alg$. Then we have
\eq{
\frac{1}{t}\lbrace V,W\rbrace^* =\; \langle V,W\rangle -t\langle\langle V,W\rangle \rangle + \mathcal{O}(t^2)\;,
}
i.e. the graded star--commutator gives the deformation of the inner product $\langle \cdot,\cdot\rangle$ up to second order. 
\end{thm}
For convenience of notation, we always denote the generalized vectors $V,W$ and their lifts to $\alg$ by the same letters. It is clear from the context which objects are used. 

\subsection{Deformation of the C-bracket}
Using theorem \ref{thm1}, we are now able to compute corrections to the C-bracket. First, it is easy to see that the Poisson structure \eqref{specialPoisson} together with the sign rule given in \eqref{supercomm} correctly reproduce the Dorfmann product $\circ$:
\eq{
V\circ W =\; \Bigl\lbrace \lbrace \theta, V\rbrace_{(1)}, W\Bigr\rbrace_{(1)}\;.
}
To see which Poisson brackets contribute to the first non-trivial corrections to $V\circ W$, we expand the double Poisson bracket up to order $t^4$:
\eq{
\Bigl\lbrace \lbrace \theta, V\rbrace,W\Bigr\rbrace^* = \;&t^2 \, V\circ W + t^3\Bigl\lbrace\lbrace \theta, V\rbrace_{(2)},W\Bigl\rbrace_{(1)} \\
+ &t^3\Bigl\lbrace\lbrace \theta, V\rbrace_{(1)}, W\Bigr\rbrace_{(2)} + \mathcal{O}(t^4)\;.
}
A short calculation shows the vanishing of $\lbrace \theta, V\rbrace_{(2)}$ for the chosen setup ($\pi$ constant) and we have
\eq{
\lbrace \theta, V\rbrace_{(1)} =\; &\xi^m\xi^n \partial_m V_n + \xi^*_k \xi^m \pi^{kn} \partial_n V_m + \xi^*_k \xi^*_m \pi^{kn}\partial_n V^m \nonumber \\
&+V_n \pi^{nm} x^*_m + x^*_n V^n\;.
}
Inserting this expression into $\Bigl\lbrace \lbrace \theta, V\rbrace_{(1)}, W\Bigr\rbrace_{(2)}$ gives exactly the contribution which was encountered for this setup in DFT, see equation \eqref{cbracketcorr}. Thus we state the following result:
\begin{thm}
Let $V = V^i\xi^*_i + V_i \xi^i$ and $W=W^i\xi^*_i + W_i \xi^i$ be two generalized vectors lifted to $\alg$, then we have
\eq{
\frac{1}{2t^2}\Bigl(\Bigl\lbrace\lbrace \theta, V\rbrace^*,W\Bigr\rbrace^* - \Bigl\lbrace\lbrace \theta, W\rbrace^*,V\Bigr\rbrace^*\Bigr) =\; [V,W]_C + t[[V,W]]_C + \mathcal{O}(t^2)\;,
}
i.e. the two-fold star commutator coincides with the $\alpha'$-corrected C-bracket of DFT up to second order in the deformation parameter $t=\alpha'$.
\end{thm}
The proof is a straight forward but lengthy evaluation in local coordinates. We refer the reader to the original article \cite{Deser:2014wva} for details, especially concerning the Koszul signs. To sum up, in the framework chosen above, it is possible to explain $\alpha'$-corrections to the bilinear pairing and C-bracket encountered in string theory via a star commutator with respect to a graded version of the Moyal-Weyl product.

\section{Outlook: $B$-, $\beta$- transformations and the Atiyah algebra}
In the final section we want to give additional evidence for the relevance of the introduced mathematical framework in physics, especially to the structures arising in DFT. First we recall that a \emph{B-transform} of a generalized vector $(X,\omega)$ is defined by
\eq{
(X,\omega)\mapsto (X, \omega + \iota_X B)\;, \quad B\in \Gamma(\wedge^2 T^*M)\;.
}
Furthermore, a $\beta$-transform is given in an analogous way by
\eq{
(X,\omega) \mapsto (X+\iota_\omega \beta, \omega)\;, \quad \beta \in \Gamma(\wedge^2 TM)\;.
}
Finally a linear transformation is given by the following definition
\eq{
(X,\omega) \mapsto (X + C(X),\omega + C^{-t}(\omega))\;, \quad C\in \Gamma(TM\otimes T^*M)\;,
}
where $A^{-t}$ means the inverse transpose of the invertible matrix $C$. The idea to lift these transformations to $\alg$ lies at hand, thus introducing the lifts 
\eq{
B = \,\tfrac{1}{2}B_{ij}\xi^i\xi^j\;,\quad \beta =\,\tfrac{1}{2}\beta^{ij}\xi^*_i\xi^*_j\;,\quad C=\,C^j_i\xi^*_j\xi^i\;,
}
it is a straight forward exercise to show that the action of $B$-,$\beta$- and linear transformations on the lift $\Sigma = X^i\xi^*_i + \omega_i \xi^i$ of a generalized vector $(X,\omega)$ is given by
\eq{
\label{Atyiah}
\Sigma \mapsto \Sigma + \lbrace&\Sigma,B\rbrace\;,\quad \Sigma \mapsto \Sigma + \lbrace \Sigma, \beta\rbrace \\
 &\Sigma \mapsto \Sigma + \lbrace \Sigma, C\rbrace\;.
}
Comparing with \cite{Roytenberg:2001am}, we see that the transformations \eqref{Atyiah} are the lifts to $\alg$ of the generators of the \emph{Atiyah algebra} of infinitesimal bundle transformations of $A\oplus A^*$, preserving the bilinear form $\eta$. With this very convenient rewriting of the transformations used frequently in the generalized geometry applications to string theory, an immediate open question is about the deformation of these transformations. The tools established in this work will be helpful to investigate this further. In addition to that, the inclusion of fluxes as ``fibre translations'' in the sense of \cite{Roytenberg:2001am} could be performed conveniently as suggested in \cite{Deser:2014wva}.


\subsection*{Acknowledgement}
I want to thank Jim Stasheff for collaboration and Athanasios Chatzistavra\-kidis, Larisa Jonke, Tom Lada, Erik Plauschinn, Dmitry Roytenberg and Theodore Voronov for discussion. Furthermore, I want to thank the organisers of the Bialowieza workshop, especially Tomasz Goli\'{n}ski and Aneta Sli\.{z}ewska for taking care especially of the newcomers to the conference venue.

\end{document}